\documentclass[10pt]{article}
\usepackage{mathrsfs}
\usepackage{subfigure}
\usepackage{graphicx}
\usepackage{amsmath}
\usepackage{amssymb}
\usepackage{caption2}
\usepackage{tabularx}
\usepackage{dcolumn}
\usepackage{array}
\usepackage{setspace}
\usepackage{booktabs}
\usepackage[figuresright]{rotating}
\begin{document}
\bibliographystyle{prsty}
\begin{center}
{\large {\bf \sc{  Could $\bar{\Lambda}_c$ and $\Lambda_c$ form bound hadronic molecule with explicit $P-wave$ ? }}} \\[2mm]
Xiu-Wu Wang$^{*}$,
Zhi-Gang  Wang$^*$\footnote{E-mail: zgwang@aliyun.com.  } \\
 Department of Physics, North China Electric Power University, Baoding 071003, P. R. China$^*$
\end{center}

\begin{abstract}
In this article, the explicit $P-wave$ of the $\bar{\Lambda}_c\Lambda_c$ dibaryon states with the $J^{PC}=0^{-\pm}$ and $1^{-\pm}$ are studied via twelve interpolated currents. Results show that the positive and negative charge conjugations for the studied states do not have too much difference, the contribution of the gluon from covariant derivative of the current with $P-wave$  could be neglected. The numerical results of the masses of the $P-wave$  $\bar{\Lambda}_c\Lambda_c$ are high above the threshold of the $\bar{\Lambda}_c$ and $\Lambda_c$ constituents, $\bar{\Lambda}_c$ and $\Lambda_c$ are unlikely to form the bound molecules with explicit $P-wave$ . Pole residues of the related states are also calculated.
\end{abstract}



\section{Introduction}
In 1935, the $\pi-meson$ exchange theory was proposed by Yukawa to explain the strong nuclear force and hence started the research era of the strong interaction \cite{Yukawa}. In the framework of quark model (QM) established in 1950s \cite{Gellmann,Zweig}, the Quantum chromodynamics (QCD) is now the fundamental theory to study the interactions of the colored quarks, gluons and the formation of hadronic matter. In the QM, the meson and baryon are made of quark-antiquark pair $q_1\bar{q}_2$ and three quarks $q_1q_2q_3$, respectively. Ever since the discovery of $X(3872)$ \cite{Choi}, many such exotic $X/Y/Z$ states interpreted as the tetraquark, pentaquark and hexaquark candidates were observed in recently years, one can check the Refs. \cite{LHCB,wangzg-zs}. There are two main kinds of interpretations for these exotic states, the hadronic molecules and compact tetraquark, pentaquark and hexaquark structures. Since the masses of these exotic states are close to the known particle-pair thresholds, many of these exotic states are the possible hadronic molecules \cite{Oset,Richard,Guo1,Martinez}. For the argument of the compact structures, the exotic states are not due to the interaction of the meson-meson, meson-baryon or baryon-baryon pairs, such as the $P_c(4312)$, $P_c(4380)$, $P_c(4440)$ and $P_c(4457)$ observed by the LHCb \cite{RAaij1,RAaij2}, a typical explanation is that they are the compact diquark-diquark-antiquark or diquark-triquark type hidden-charm pentaquark states \cite{compact1,compact2,compact3,compact4,compact5,compact6,compact7}.

In 1980s, the topic of dibaryon and baryonium which are due to the interaction of two baryons had been caused great attention both theoretically and experimentally \cite{Strakovsky}. Till now, the universally accepted dibaryon state determined by experiments is the deuteron which is composed by a proton and neutron \cite{Harold}. Another typical example is the $d^*(2380)$, it is recognized as the hadronic molecule of two $\Delta$ baryons \cite{Bashkanov, Adlarson1, Adlarson2}. However, debate is there for the $d^*(2380)$, the `dibaryon $d^*(2380)$ peak' is interpreted as the triangle singularity effect in Refs. \cite{Oset1,Oset2}. In recent years, several QCD-inspired models are applied and lots of predictions have been made for the dibaryons and baryoniums waiting for the testification of experimental data, such as the quark delocalization color screening model \cite{Xia,Huang1}, solving the Bethe-Salpeter equation via effective Lagrangians \cite{Lu,Zhu1}, constituent quark model \cite{Huang2,Carames}, lattice QCD \cite{Morita1,Morita2}, the QCD sum rules \cite{Kodama,Chen,Wang2,Wan,SWZG1,SWZG2,Qiao1,Qiao22} and so on. In 2018, the BESIII Collaboration reported the nearly flat cross section behavior for the $e^+e^-\rightarrow \Lambda_c\bar{\Lambda}_c$ precess just above the $\Lambda_c\bar{\Lambda}_c$ threshold \cite{BESIII-dilamtc}. In Ref. \cite{BESIII-dilamtc11}, the BESIII Collaboration studied the process $e^+e^-\rightarrow \Lambda_c\bar{\Lambda}_c$ again at twelve center-of-mass energies from 4.6119 to 4.9509 GeV, and pointed out that flat cross sections around 4.63 GeV are obtained and no indication of the resonant structure $Y(4630)$, as reported by Belle, is found. In Refs. \cite{Salnikov1,Salnikov2}, the authors analyzed the BESIII experimental data. The group argued that there is a bound state of $\Lambda_c\bar{\Lambda}_c$ \cite{Salnikov1}. And now, many theoretic groups also argue that there are bound $\Lambda_c\bar{\Lambda}_c$ baryonium states \cite{Lee-Lambt-baryonium,RChen-Lambt-baryonium,DSong-Lambt-baryonium,JXLu-Lambt-baryonium,YDChen-Lambt-baryonium,XKDong-Lambt-baryonium,SJCao-Lambt-baryonium}, but for our study in Ref. \cite{XWWang-Lambt-baryonium}, no bound state has ever been found for the $J^P=0^+,\,\,0^-,\,\,1^+$ and $1^-$, similar conclusion is also proposed in Ref. \cite{Wan}. Thus, whether the $\Lambda_c$ and $\bar{\Lambda}_c$ could form the hadronic molecule or not is a topic under debate. Inspired by this, the question such that could the $\Lambda_c$ and $\bar{\Lambda}_c$ baryon form the bound states with $P-wave$  is put forward, and QCD sum rules is applied for the solution of this problem in the present study. If the theory could predict some physical conclusions in this topic which could be testified by the experiment in the future, it will strengthen the trueness of the interaction of the $\Lambda_c$ and $\bar{\Lambda}_c$ studied by the theory in return.

The QCD sum rules is a non-perturbative theory proposed by Shifman, Vainshtein and Zakharov in 1979 \cite{Shifman1,Shifman2}. Its main technique is the consideration of the perturbative and vacuum condensates contributions via the performance of operator product expansion of the full propagators. It has been extensively applied to study the hidden-charm (bottom)
tetraquark (molecular) states \cite{xzAgaev,xzChen1,xzChen2,xzOzdem,xzLee1,xzZhang1,xzZhang2,xzMatheus,xzMihara,xzAlbuquerque,xzDi,Wang3,Wang4,Wang5}, pentaquark (molecular) states \cite{Wang66,Wang77} and hexaquark (molecular) states \cite{WZG}. Especially, In Ref. \cite{WZG}, the two-baryon scattering states are considered, however, their contributions can be neglected. In our previous works, the $\Lambda_c\Lambda_c$, $\Sigma_c\Sigma_c$ dibaryons and $\Lambda_c\bar{\Lambda}_c$, $\Sigma_c\bar{\Sigma}_c$ baryoniums have already been studied \cite{XWWang-Lambt-baryonium,XWWang-disigma}, however, the systematic study for the hidden-charm dibaryons and baryoniums is still far beyond completion.

The article is organized as follows: in Sect. 2, the QCD sum rules for the dibaryon states with $P-wave$ are derived; the numerical results and discussions are given in Sect. 3; Sect. 4 is reserved for the conclusions.

\section{QCD sum rules for the dibaryon states with $P-wave$}
The interpolating currents applied to study the $\Lambda_c$ baryon with the $J^P=\frac{1}{2}^+$ in the ground state and $\Lambda_c$ baryon with the $J^P=\frac{1}{2}^-$ and $P-wave$  are marked as $J_0(x)$ and $J_1(x)$, moreover, the two possible currents to interpolate $\Lambda_c$ baryon with the $J^P=\frac{3}{2}^-$ and $P-wave$  are represented as $J_{2,\mu}(x)$ and $J_{3,\mu}(x)$, respectively. They are expressed as \cite{Qxin},
\begin{eqnarray}
\notag J_0(x) &=& \varepsilon^{ijk}\left[u^{iT}(x)C\gamma_5d^j(x)\right]c^k(x)\, ,\\
\notag J_1(x) &=& \varepsilon^{ijk}\left[\partial^{\mu}u^{iT}(x)C\gamma^{\nu}d^j(x)-u^{iT}(x)C\gamma^{\nu}\partial^{\mu}d^j(x)\right]\sigma_{\mu\nu}c^k(x)\, ,\\
\notag J_{2,\mu}(x) &=& \varepsilon^{ijk}\left[\partial^{\alpha}u^{iT}(x)C\gamma^{\beta}d^j(x)-u^{iT}(x)C\gamma^{\beta}\partial^{\alpha}d^j(x)\right](\widetilde{g}_{\mu\alpha}\gamma_{\beta}-\widetilde{g}_{\mu\beta}\gamma_{\alpha})\texttt{i}\gamma_5c^{k}(x)\,,\\
\notag J_{3,\mu}(x) &=& \varepsilon^{ijk}\left[\partial^{\alpha}u^{iT}(x)C\gamma^{\beta}d^j(x)-u^{iT}(x)C\gamma^{\beta}\partial^{\alpha}d^j(x)\right]\\
 &&\cdot\left(g_{\mu\alpha}\gamma_{\beta}+g_{\mu\beta}\gamma_{\alpha}-\frac{1}{2}g_{\alpha\beta}\gamma_{\mu}\right)\texttt{i}\gamma_5c^{k}(x)\,,
\end{eqnarray}
where, $\texttt{i}^2=-1$, $i$, $j$ and $k$ are the color indices, $\varepsilon^{ijk}$ represents the antisymmetric tensor in the color space, $C$ stands for the charge conjugation matrix, $\alpha$, $\beta$, $\mu$ and $\nu$ are the Lorentz indices, the superscript $T$ means the  transpose of the related quark spinor in the Dirac spinor space, $\sigma_{\mu\nu}=\frac{\texttt{i}}{2}(\gamma_{\mu}\gamma_{\nu}-\gamma_{\nu}\gamma_{\mu})$,  $\widetilde{g}_{\mu\alpha}=g_{\mu\alpha}-\frac{1}{4}\gamma_{\mu}\gamma_{\alpha}$, $\partial_{\mu}$ is the partial derivative with respect to $x^{\mu}$, since only the covariant derivative $D_{\mu}=\partial_{\mu}-ig_sG_{\mu}$ is gauge invariant, the following currents with $P-wave$  interpolated via the covariant derivatives are also considered,
\begin{eqnarray}
\notag \eta_1(x) &=& \varepsilon^{ijk}\left[D^{\mu}u^{iT}(x)C\gamma^{\nu}d^j(x)-u^{iT}(x)C\gamma^{\nu}D^{\mu}d^j(x)\right]\sigma_{\mu\nu}c^k(x)\, ,\\
\notag \eta_{2,\mu}(x) &=& \varepsilon^{ijk}\left[D^{\alpha}u^{iT}(x)C\gamma^{\beta}d^j(x)-u^{iT}(x)C\gamma^{\beta}D^{\alpha}d^j(x)\right](\widetilde{g}_{\mu\alpha}\gamma_{\beta}-\widetilde{g}_{\mu\beta}\gamma_{\alpha})\texttt{i}\gamma_5c^{k}(x)\,,\\
\notag \eta_{3,\mu}(x) &=& \varepsilon^{ijk}\left[D^{\alpha}u^{iT}(x)C\gamma^{\beta}d^j(x)-u^{iT}(x)C\gamma^{\beta}D^{\alpha}d^j(x)\right]\\
 &&\cdot\left(g_{\mu\alpha}\gamma_{\beta}+g_{\mu\beta}\gamma_{\alpha}-\frac{1}{2}g_{\alpha\beta}\gamma_{\mu}\right)\texttt{i}\gamma_5c^{k}(x)\,.
\end{eqnarray}
The difference between the partial derivative and covariant derivative lies in the additional contribution of the gluon from the covariant derivative, applying the Fock-Schwinger gauge, the gluon field is expressed as,
\begin{eqnarray}
G^n_{\mu}&=&\frac{1}{2\cdot 0!}x^{\alpha}G_{\alpha\mu}^n(0)+\frac{1}{3\cdot 1!}x^{\alpha}x^{\beta}D_{\beta}G^n_{\alpha\mu}(0)+\cdot\cdot\cdot\,\,,
\end{eqnarray}
where, $G_{\mu}=G^n_{\mu}t^n$, $t^n=\frac{\lambda^n}{2}$, $\lambda^n$ are the Gell-Mann matrices ($n=1,2,\cdot\cdot\cdot,8$). The first-order approximation is chosen for the QCD sum rules, namely, the contribution of $x^\alpha x^\beta$ and other high-order terms are neglected for the gluon field.

Under parity transformation $\widehat{P}$ and operation of charge conjugation $\widehat{C}$, the above currents satisfy the following equations,
\begin{eqnarray}
\notag &&\widehat{P}J_0(x)\widehat{P}^{-1}=\gamma^0J_0(\widetilde{x}) \, ,\\
\notag &&\widehat{P}J(x)\widehat{P}^{-1}=-\gamma^0J(\widetilde{x})\, , \\
&&\widehat{P}J_{\mu}(x)\widehat{P}^{-1}=\gamma^0J^\mu(\widetilde{x})\,,
\end{eqnarray}
\begin{eqnarray}
\notag &&\widehat{C}J_0(x)\widehat{C}^{-1}=-C\bar{J}_0^T(x)\, , \\
\notag &&\widehat{C}J(x)\widehat{C}^{-1}=-C\bar{J}^T(x)\, , \\
 &&\widehat{C}J_{\mu}(x)\widehat{C}^{-1}=-C\bar{J}^T_{\mu}(x)\, , 
\end{eqnarray}
where, $\widetilde{x}=(x^0,-x^1,-x^2,-x^3)$, and
\begin{eqnarray}
\notag J(x)&=&J_1(x),\,\,\eta_1(x)\\
\notag J_{\mu}(x)&=&J_{2,\mu}(x),\,\,J_{3,\mu}(x),\,\,\eta_{2,\mu}(x),\,\,\eta_{3,\mu}(x)\,.
\end{eqnarray}
Based on the above equations, the following twelve currents of parity and charge conjugation eigenstates  are constructed to couple the $\bar{\Lambda}_c\Lambda_c$ with $P-wave$ ,
\begin{eqnarray}
\notag \mathcal{J}^{\pm}_1 &=& \bar{J}_0J_1\pm\bar{J}_1J_0\, , \\
\notag \mathcal{J}^{\pm}_{2,\mu} &=& \bar{J}_0J_{2,\mu}\pm\bar{J}_{2,\mu}J_0\, , \\
\notag \mathcal{J}^{\pm}_{3,\mu} &=& \bar{J}_0J_{3,\mu}\pm\bar{J}_{3,\mu}J_0\, , \\
\notag \mathcal{J}^{\pm}_4 &=& \bar{J}_0\eta_1\pm\bar{\eta}_1J_0\, , \\
\notag \mathcal{J}^{\pm}_{5,\mu} &=& \bar{J}_0\eta_{2,\mu}\pm\bar{\eta}_{2,\mu}J_0\, , \\
 \mathcal{J}^{\pm}_{6,\mu} &=& \bar{J}_0\eta_{3,\mu}\pm\bar{\eta}_{3,\mu}J_0\, ,
\end{eqnarray}
one can testify that all these twelve currents have the negative parity and $\widehat{C}\mathcal{J}^{\pm}\widehat{C}^{-1}=\pm\mathcal{J}^{\pm}$, where, the subscripts of the currents are omitted.

Henceforward, the correlation functions are written as,
\begin{eqnarray}
\notag \Pi(p)&=&\texttt{i}\int d^4x e^{\texttt{i}p\cdot x}\langle 0 | \textsc{T}\{\mathcal{J} (x) \mathcal{J}^{\dag}(0) \}| 0\rangle \,,\\
\Pi_{\mu\nu}(p)&=&\texttt{i}\int d^4x e^{\texttt{i}p\cdot x}\langle 0 |\textsc{T}\{ \mathcal{J}_{\mu} (x) \mathcal{J}_{\nu}^{\dag}(0)\} | 0\rangle \, ,
\end{eqnarray}
where, the superscript $\pm$ for $\Pi(p)$, $\Pi_{\mu\nu}(p)$, $\mathcal{J}(x)$ and $\mathcal{J}_{\mu}(x)$ is neglected, $\textsc{T}$ represents the time-ordering operation, and
\begin{eqnarray}
\notag \mathcal{J}(x) &=& \mathcal{J}_1^{\pm}(x),\,\,\mathcal{J}_4^{\pm}(x) \,,\\
\notag \mathcal{J}_{\mu}(x) &=& \mathcal{J}_{2,\mu}^{\pm}(x),\,\, \mathcal{J}_{3,\mu}^{\pm}(x),\,\, \mathcal{J}_{5,\mu}^{\pm}(x),\,\, \mathcal{J}_{6,\mu}^{\pm}(x)\,.
\end{eqnarray}
The conventional routine is followed as to insert a complete set of intermediate hadronic states with the same quantum numbers as the currents $\mathcal{J}(x)$ and $\mathcal{J}_{\mu}(x)$ for the hadronic sides of the correlation functions $\Pi(p)$ and $\Pi_{\mu\nu}(p)$, the contribution of the ground states are separated, thus the correlation functions at the hadronic sides are phenomenologically derived as,
\begin{eqnarray}
\notag && \Pi(p)=\frac{\lambda^2_{\mathcal{Z}}}{m_{\mathcal{Z}}^2-p^2}+\cdot\cdot\cdot\, , \\
&&\Pi_{\mu\nu}(p)=\frac{\lambda^2_{\mathcal{Z}}}{m_{\mathcal{Z}}^2-p^2} \left(-g_{\mu\nu}+\frac{p_\mu p_\nu}{p^2}\right)+\cdot\cdot\cdot \, ,
\end{eqnarray}
where,
\begin{eqnarray}
\notag && \langle0 | \mathcal{J}(0)|\mathcal{Z}(p) \rangle =\lambda_{\mathcal{Z}}\, ,\\
&&\langle0 | \mathcal{J}_{\mu}(0)|\mathcal{Z}(p) \rangle =\lambda_{\mathcal{Z}}\epsilon_\mu \, ,
\end{eqnarray}
in the above expressions, $|\mathcal{Z}\rangle$ denote the twelve ground states interpolated by the related currents, $\epsilon_\mu$ are the polarization vectors of the corresponding states interpolated by $\mathcal{J}_{\mu}(x)$.

At the QCD sides, Wick theorem is applied to contract the quark fields and the correlation functions in terms of full quark propagators are derived. Conduct the operator product expansion, the full quark propagators are shown as,
\begin{eqnarray}
\notag\ Q^{ab}(x)&=& \frac{\texttt{i}x\!\!\!/\delta^{ab}}{2\pi^{2}x^{4}}-\frac{\delta^{ab}}{12}\langle\overline{q}q\rangle-\frac{\delta^{ab}x^2}{192}\langle\overline{q}g_s\sigma G q\rangle-\frac{\texttt{i}\delta^{ab}x^2x\!\!\!/g_s^2\langle\overline{q}q\rangle^2}{7776}\\
\notag &&-\left(t^n\right)^{ab}\left(x\!\!\!/\sigma^{\alpha\beta}+\sigma^{\alpha\beta}x\!\!\!/\right)\frac{\texttt{i}}{32\pi^2x^2}g_s G_{\alpha\beta}^n  \\
\notag\ &&-\frac{\delta^{ab}x^4\langle\overline{q}q\rangle\langle GG \rangle}{27648}-\frac{1}{8}\langle\overline{q}^b\sigma^{\alpha\beta}q^a\rangle\sigma_{\alpha\beta}-\frac{1}{4}\langle\overline{q}^b\gamma_\mu q^a\rangle\gamma^\mu+\cdot\cdot\cdot\, ,
\end{eqnarray}
\begin{eqnarray}
\notag\
B_{ab}(x)&=&\frac{\texttt{i}}{(2\pi)^{4}}\int d^{4}ke^{-\texttt{i}k\cdot x}\bigg\{\frac{\delta_{ab}}{k\!\!\!/-m_{c}}-\frac{g_{s}G_{\alpha\beta }^{h}t_{ab}^{h}}{4}\frac{\sigma^{\alpha\beta}(k\!\!\!/+m_{c})+(k\!\!\!/+m_{c})\sigma ^{\alpha
\beta }}{(k^{2}-m_{c}^{2})^{2}}\\
\notag\
&&+\frac{g_{s}D_{\alpha}G_{\beta\lambda}^{h}t_{ab}^{h}\left(f^{\lambda\beta\alpha}+f^{\lambda\alpha\beta}\right)}{3(k^{2}-m_{c}^{2})^{4}}
+\cdot\cdot\cdot\bigg \}\, ,
\end{eqnarray}
\begin{eqnarray}
 &&f^{\lambda \alpha \beta }=(k\!\!\!/+m_{c})\gamma ^{\lambda
}(k\!\!\!/+m_{c})\gamma ^{\alpha }(k\!\!\!/+m_{c})\gamma ^{\beta
}(k\!\!\!/+m_{c})\, ,
\end{eqnarray}
where, $Q^{ab}(x)$ and $B_{ab}(x)$ are the light and heavy quark propagators, respectively, $a$ and $b$ are the color indices, $m_c$ is the mass of the $c$ quark.

For the chosen of vacuum condensates, the correlation functions of the considered states contain two heavy and four light quark lines in a consistent way. Consider the heavy quark line emits a gluon and each light quark line contributes quark-antiquark pair, they form the quark-gluon operator $g_sG_{\alpha\beta}g_sG_{\eta\tau}\overline{q}q\overline{q}q\overline{q}q\overline{q}q$ with dimension 16, the operator leads to the vacuum condensates $\langle\frac{\alpha_s}{\pi}GG\rangle\langle\overline{q}q\rangle^4$ and $\langle\overline{q}g_s\sigma Gq\rangle^2\langle\overline{q}q\rangle^2$, for this sake, the highest dimension of the vacuum condensates taking into consideration in the present study is 16. For the truncation of the order $\mathcal{O}(\alpha_s^k )$, it is accurate enough to calculate terms for $k\leq1$ \cite{XWWang-Lambt-baryonium}. Above all, the selected vacuum condensates after the operator product expansion are $\langle\frac{\alpha_s}{\pi}GG\rangle$, $\langle\overline{q}q\rangle^2$, $\langle\overline{q}g_s\sigma Gq\rangle\langle\overline{q}q\rangle$, $\langle\overline{q}g_s\sigma Gq\rangle^2$, $\langle\frac{\alpha_s}{\pi}GG\rangle\langle\overline{q}q\rangle^2$, $g_s^2\langle\overline{q}q\rangle^4$, $\langle\overline{q}g_s\sigma Gq\rangle\langle\overline{q}q\rangle^3$, $\langle\overline{q}g_s\sigma Gq\rangle^2\langle\overline{q}q\rangle^2$ and $\langle\frac{\alpha_s}{\pi}GG\rangle\langle\overline{q}q\rangle^4$.

For the correlation functions, the integrals of the light and heavy quarks are taken in the space and linear momentum coordinates, respectively. Since the masses of the light quarks are too small to make much difference, they are neglected in the present calculation. Conduct the Borel transformation for both the hadronic and QCD sides and apply the quark-hadron dual, the QCD sum rules are derived as,
\begin{eqnarray}
\notag \lambda_{\mathcal{Z}}^2\exp\left(-\frac{m_{\mathcal{Z}}^2}{T^2}\right)&=&\int_{\Delta^2}^{s_0}ds \rho_{\mathcal{Z},{\rm QCD}}(s)\exp\left(-\frac{s}{T^2}\right)\, ,\\
m_{\mathcal{Z}}^2&=&\frac{-\frac{d}{d\tau}\int_{\Delta^2}^{s_0}ds \rho_{\mathcal{Z},{\rm QCD}}(s)\exp(-s\tau)}{\int_{\Delta^2}^{s_0}ds \rho_{\mathcal{Z},{\rm QCD}}(s)\exp(-s\tau)}\, ,
\end{eqnarray}
where, the superscript $\pm$ for $\lambda_{\mathcal{Z}}$, $m_{\mathcal{Z}}$ and $\rho_{\mathcal{Z},{\rm QCD}}$ is neglected, $\Delta^2=4m_c^2$, $\tau=\frac{1}{T^2}$, $s_0$ is the continuum threshold parameter of the related state, its experiential value is determined as $\sqrt{s_0}=m_{\mathcal{Z}}+0.5\sim0.7\,\,\rm{GeV}$ \cite{XWWang-Lambt-baryonium,XWWang-disigma}.

Taking the spectral density of state coupled by current $\mathcal{J}_1^+$ for example, its analytical expression is attached in the Appendix. Compare the analytical expressions of the spectral densities of the positive and negative charge eigenstates, the only differences come from the contribution of the vacuum condensation $\langle\overline{q}g_s\sigma Gq\rangle^2$. As for the differences of the spectral densities between the partial derivative and covariant derivative, the additional contribution of vacuum condensates due to the gluon from the covariant derivative also lie in the vacuum condensate $\langle\overline{q}g_s\sigma Gq\rangle^2$.
\section{Numerical results and discussions}
The standard values of the vacuum condensates are applied, in detail, $\langle\overline{q}q\rangle=-(0.24\pm0.01\;{\rm GeV})^3$, $\langle\overline{q}g_s\sigma Gq\rangle=m_0^2\langle\overline{q}q\rangle$, $m_0^2=(0.8\pm0.1)\;{\rm GeV}^2$, $\langle\frac{\alpha_s}{\pi}GG\rangle=(0.012\pm0.004)\;{\rm GeV}^4$ at the energy scale $\mu=1\;{\rm GeV}$ \cite{Shifman1,Reinders,ColangeloReview}, $\overline{MS}$ mass $m_c(m_c)=(1.275\pm0.025)\;{\rm GeV}$ \cite{PDG}. For these input parameters, their energy-scale dependence are,
\begin{eqnarray}
\notag \langle\overline{q}q\rangle(\mu)&=&\;\;\langle\overline{q}q\rangle(1{\rm GeV})\left[\frac{\alpha_s(1{\rm GeV})}{\alpha_s(\mu)}\right]^{\frac{12}{33-2n_f}}\, ,\\
\notag \langle\overline{q}g_s\sigma Gq\rangle(\mu)& = &\;\;\langle\overline{q}g_s\sigma Gq\rangle(1{\rm GeV})\left[\frac{\alpha_s(1{\rm GeV})}{\alpha_s(\mu)}\right]^{\frac{2}{33-2n_f}}\, ,\\
\notag  m_c(\mu)&=&\;\;m_c(m_c)\left[\frac{\alpha_s(\mu)}{\alpha_s(m_c)}\right]^{\frac{12}{33-2n_f}}\, ,\\
\notag \alpha_s(\mu)&=&\;\;\frac{1}{b_0t}\left[1-\frac{b_1}{b_0^2}\frac{\rm{log}\emph{t}}{t}+\frac{b_1^2(\rm{log}^2\emph{t}-\rm{log}\emph {t}-1)+\emph{b}_0\emph{b}_2}{b_0^4t^2}\right]\, ,
\end{eqnarray}
where, $t=\rm{log}\frac{\mu^2}{\Lambda_{\emph{{\rm QCD}}}^2}$, $\emph b_0=\frac{33-2\emph{n}_\emph{f}}{12\pi}$, $b_1=\frac{153-19n_f}{24\pi^2}$, $b_2=\frac{2857-\frac{5033}{9}n_f+\frac{325}{27}n_f^2}{128\pi^3}$
and $\Lambda_{{\rm QCD}}=213$ MeV, $296$ MeV, $339$ MeV for the flavors $n_f=5,\,4,\,3$, respectively \cite{PDG,Narison}, in the present study, the flavor number $n_f=4$. The following data are from the PDG \cite{PDG}: the mass of the $\Lambda_c$ baryon with $J^P=\frac{1}{2}^+$ is $2295\,\,\rm{MeV}$, the mass with $J^P=\frac{1}{2}^-$ and $P-wave$ is $2595\,\,\rm{MeV}$ and the mass with $J^P=\frac{3}{2}^-$ and $P-wave$  is $2625\,\,\rm{MeV}$.  The energy scale formula is applied to determine the best energy scale for the numerical calculation \cite{WZG-penta-mole-CPC,WZGNNN2}, which is given by,
\begin{eqnarray}
\mu=\sqrt{M_{X/Y/Z/P}^2-4\mathbb{M}_c^2}\, ,
\end{eqnarray}
where, $\mathbb{M}_c$ is the effective charm quark mass, and its updated value is $\mathbb{M}_c=1.85\pm0.01$ \rm GeV \cite{WZG-penta-mole-CPC}, $M_{X/Y/Z/P}$ represent the masses of the exotic states. The pole contributions (PC) is defined to judge the pole dominance criterion of the QCD sum rules, moreover, the convergence of the operator product expansion is reflected via $\textsc{D}(n)$, the contribution of vacuum condensate with dimension $n$, they are written as,
\begin{eqnarray}
\notag {\rm PC}&=&\frac{\int_{4m_c^2}^{s_0}ds\rho_{{\rm QCD}}(s)\exp\left(-\frac{s}{T^2}\right)} {\int_{4m_c^2}^{\infty}ds\rho_{{\rm QCD}}(s) \exp\left(-\frac{s}{T^2}\right)}\, ,\\
\textsc{D}(n)&=&\frac{\int_{4m_c^2}^{s_0}ds\rho_{{\rm QCD}}(s,n) \exp\left(-\frac{s}{T^2}\right)}{\int_{4m_c^2}^{s_0}ds\rho_{{\rm QCD}}(s) \exp\left(-\frac{s}{T^2}\right)}\, ,
\end{eqnarray}
where, $\rho_{{\rm QCD}}(s,n)$ is the spectral density contributed by the vacuum condensates with dimension $n$ picked out from the corresponding $\rho_{{\rm QCD}}(s)$. The dimensional contribution is normalized as $\mathcal{D}(n)=\frac{|\textsc{D}(n)|}{\sum|\textsc{D}(n)|}$. The best energy scales and continuum threshold parameters for the corresponding states are determined via trivial and trial in the numerical calculation \cite{XWWang-Lambt-baryonium}.

Numerical results show that the $m_{\mathcal{Z}}-T^2$ or the $\lambda_{\mathcal{Z}}-T^2$ curves are basically the same for the currents with partial derivatives and covariant derivatives, it is reasonable since the difference of the two kinds of spectral densities lie only in the contribution of vacuum condensates $\langle\overline{q}g_s\sigma Gq\rangle^2$ with dimension 10, whereas $\mathcal{D}(10)<3.30\%$ holds for all the states considered in the present study. Moreover, under the same input parameters, the masses with positive charge conjugation extracted from the Borel platforms only have the differences of a few $\rm{MeV}$ compared with the related negative ones, in detail, $m_1^+-m_1^-=3\,\,\rm{MeV}$, $m_2^--m_2^+=2\,\,\rm{MeV}$ and $m_3^--m_3^+=1\,\,\rm{MeV}$.  Taking the $m_1-T^2$ curves for example, the curves of positive and negative charge conjugations drawn by the central input parameters are shown in the Fig.1. For this sake, only the $m_{\mathcal{Z}}-T^2$ and $\lambda_{\mathcal{Z}}-T^2$ curves for the currents with positive charge conjugation and partial derivatives are drawn in the Fig.2. The detailed dimensional contribution of these three states are shown in the Fig.3, and the extracted masses and pole residues from the centers of the Borel platforms are listed in the Table 2. Values of the necessary parameters are presented in the Table 1.

\begin{figure}
 \centering
 \includegraphics[totalheight=5cm,width=7cm]{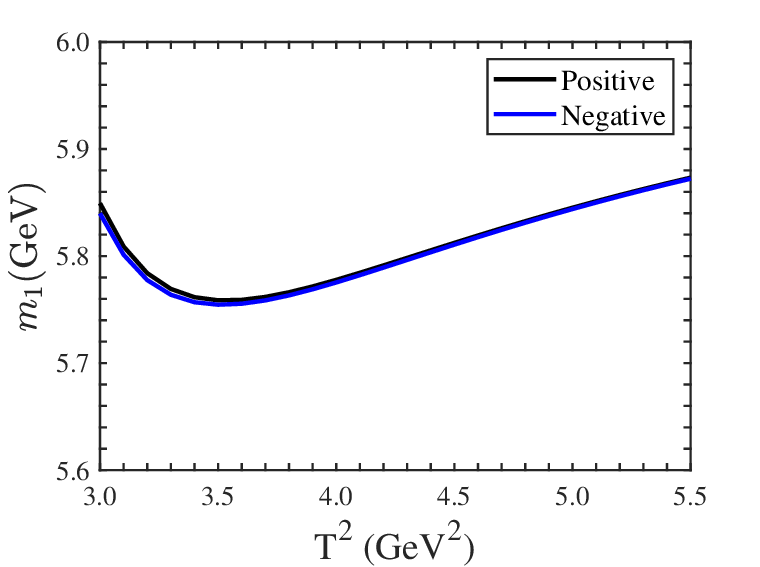}
 \caption{The $m_1-T^2$ curves plotted by the same central input parameters for both the positive and negative charge conjugations interpolated by the currents $\mathcal{J}_1^{\pm}$ .}\label{baryon-fig1}
\end{figure}

\begin{figure}
 \centering
 \includegraphics[totalheight=5cm,width=7cm]{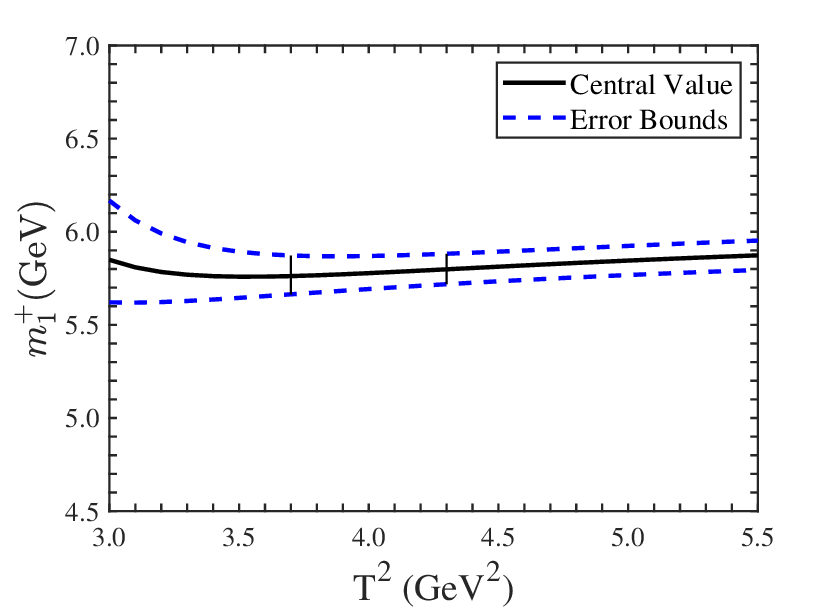}
 \includegraphics[totalheight=5cm,width=7cm]{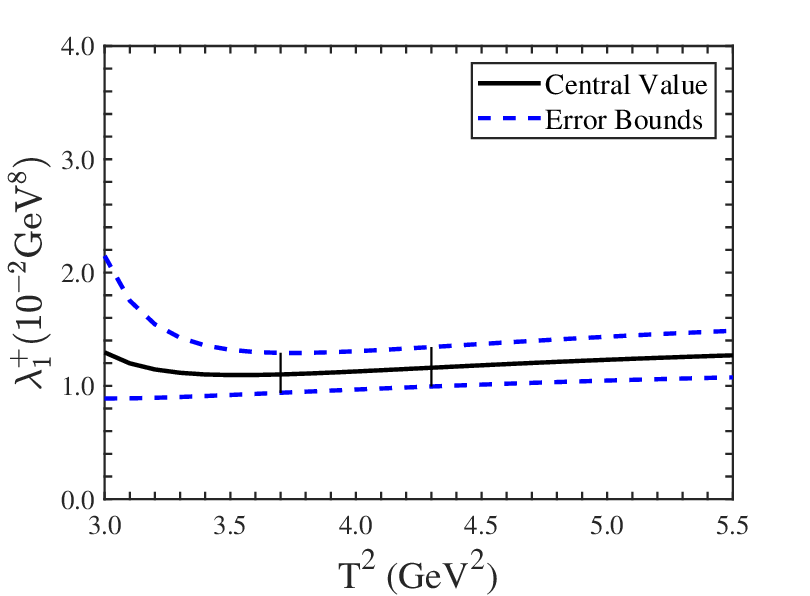}
  \includegraphics[totalheight=5cm,width=7cm]{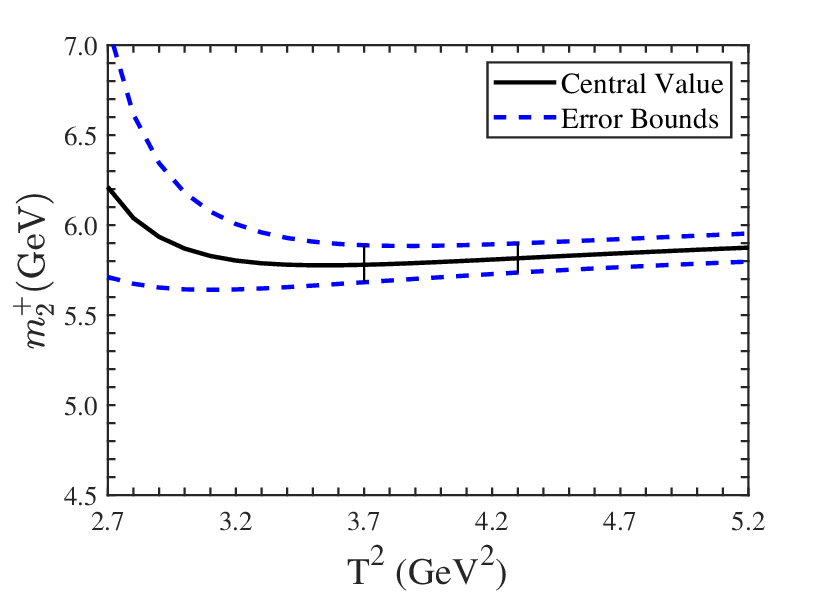}
 \includegraphics[totalheight=5cm,width=7cm]{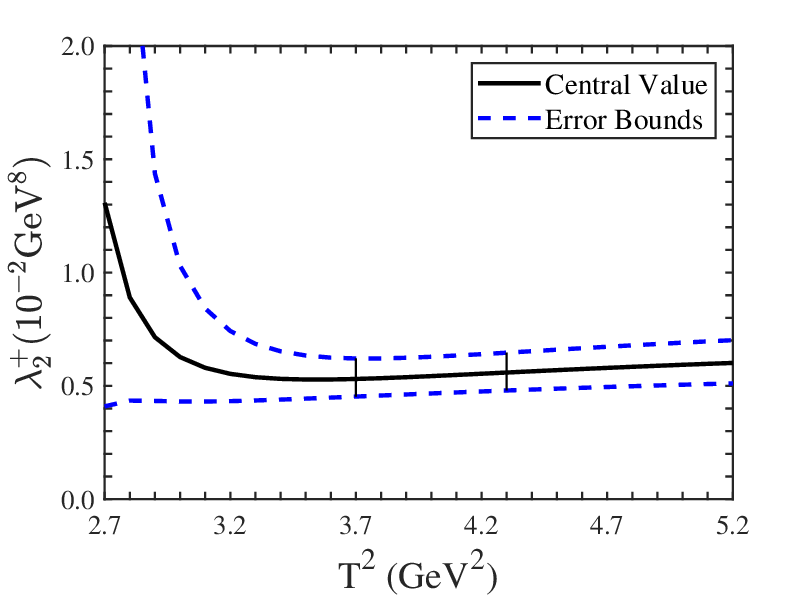}
  \includegraphics[totalheight=5cm,width=7cm]{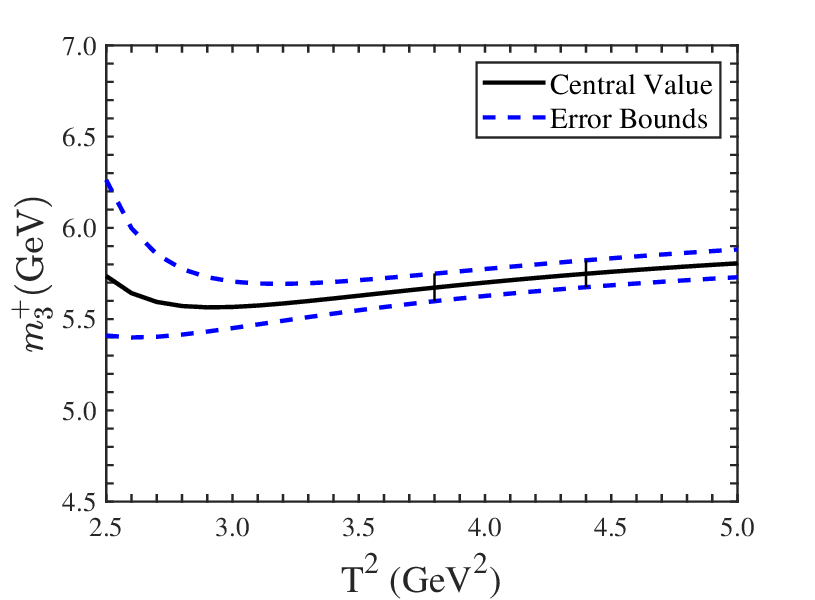}
 \includegraphics[totalheight=5cm,width=7cm]{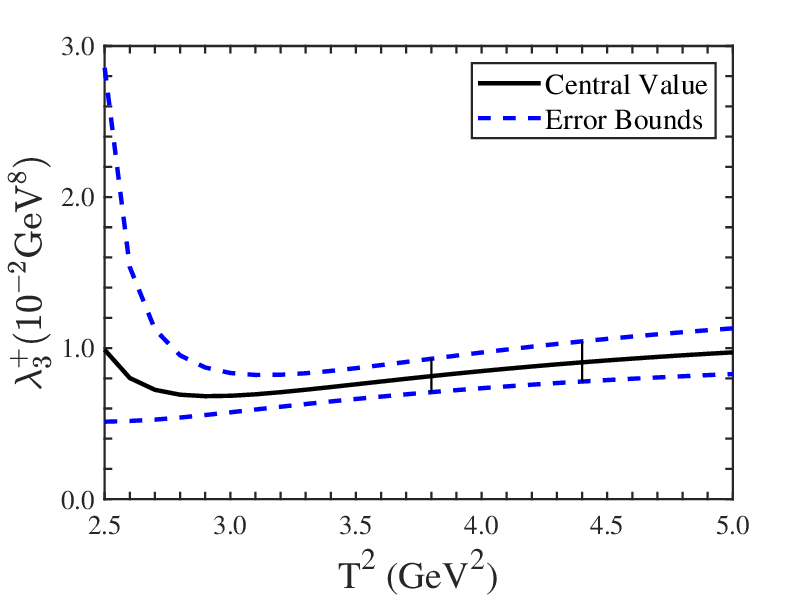}
 \caption{The masses and pole residues of the $\bar{\Lambda}_c\Lambda_c$ dibaryons with $P-wave$ .}\label{baryon-fig2}
\end{figure}

\begin{figure}
 \centering
 \includegraphics[totalheight=5cm,width=7cm]{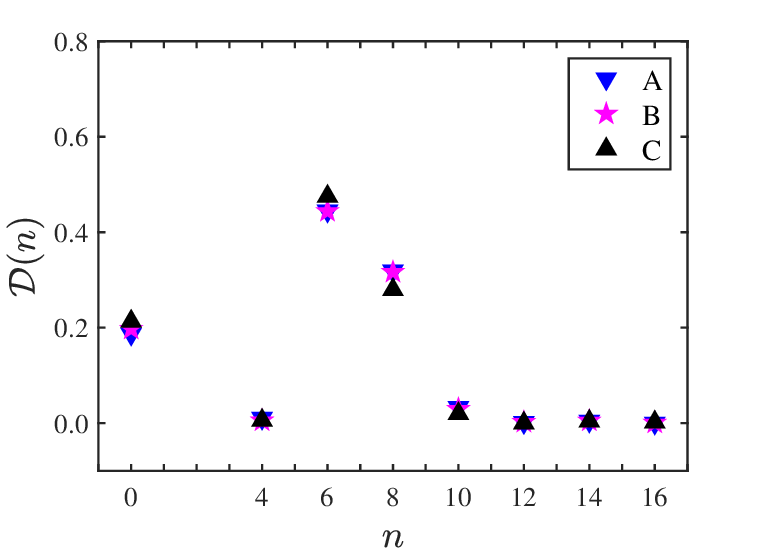}
 \caption{The dimensional contribution of the states, where, A, B and C represent the $\mathcal{D}(n)$ of the states interpolated by the currents $\mathcal{J}_1^+$, $\mathcal{J}_{2,\mu}^+$ and $\mathcal{J}_{3,\mu}^+$, respectively.}\label{baryon-fig3}
\end{figure}

For the $\bar{\Lambda}_c\Lambda_c$ dibaryon with $P-wave$  interpolated by the current $\mathcal{J}_1^+$, its mass is $m_1^+=5.78^{+0.09}_{-0.09}\,\,\rm{GeV}$ which is $888\,\,\rm{MeV}$ above the threshold of the two $\bar{\Lambda}_c$ and $\Lambda_c$ baryon constituents, even if taking the error bounds into consideration, the lower bound of its mass is still far larger than the threshold of the two baryon constituents. Thus it is unlikely for the $\bar{\Lambda}_c\Lambda_c$ dibaryon with $P-wave$  and $J^P=0^-$ to form the bound hadronic molecule, this state is assigned as the resonance dibaryon state. Judging from the the pole contribution of this state, PC ranges from $58\%$ to $41\%$ among its Borel platform, the pole dominance criterium for the QCD sum rules holds. Considering the dimensional contribution of this state, the most important contribution are from the $\mathcal{D}(0)$, $\mathcal{D}(6)$ and $\mathcal{D}(8)$ which are due to the leading order, $\langle\overline{q}q\rangle^2$ and $\langle\overline{q}g_s\sigma Gq\rangle\langle\overline{q}q\rangle$, respectively. For the contribution of the highest dimensions $\mathcal{D}(12)$, $\mathcal{D}(14)$ and $\mathcal{D}(16)$, their numerical results are $0.01\%$, $0.4\%$ and $0.2\%$, respectively, it shows that the convergency of the operator product expansion holds well. Moreover, the energy scale formula is satisfied. For $\bar{\Lambda}_c\Lambda_c$ dibaryon states with $P-wave$  interpolated by the vector currents $\mathcal{J}_{2,\mu}^+$ and $\mathcal{J}_{3,\mu}^+$, similar conclusions can be derived, they are all resonance dibaryon states.

\begin{table}
\caption{ The  optimal energy scales $\mu$, Borel windows $T^2$, continuum threshold parameters $s_0$ and
 pole contributions (PC)    for $\bar{\Lambda}_c\Lambda_c$ states with $P-wave$.} \label{Borel-pole}
\begin{center}
\begin{tabular}{|c|c|c|c|c|c|c|c|}\hline\hline
                         &$J^P$                   &$\mu(\rm GeV)$  &$T^2 (\rm{GeV}^2)$  &$\sqrt{s_0}(\rm{GeV})$   &PC     \\ \hline
$\mathcal{J}_1^+$         &$0^-$      &4.4     &$3.7-4.3$        &$6.38\pm0.10$       &$(41-58)\%$  \\   \hline

$\mathcal{J}_{2,\mu}^+$   &$1^-$      &4.5     &$3.7-4.3$        &$6.40\pm0.10$       &$(41-59)\%$  \\   \hline

$\mathcal{J}_{3,\mu}^+$   &$1^-$      &4.3     &$3.8-4.4$        &$6.38\pm0.10$       &$(40-58)\%$  \\   \hline
\hline
\end{tabular}
\end{center}
\end{table}

\begin{table}
\caption{ The predicted masses and pole residues of the $\bar{\Lambda}_c\Lambda_c$ states with $P-wave$.} \label{mass-pole-residue-tab}
\begin{center}
\begin{tabular}{|c|c|c|c|c|c|c|c|}\hline\hline
             &$J^P$   &$m (\rm{GeV})$   &$\lambda (10^{-2}\rm{GeV}^8)$ & Thresholds (MeV) & Assignments   \\  \hline
$\mathcal{J}_1^+$  &$0^-$   &$5.78_{-0.09}^{+0.09}$  &$1.13^{+0.18}_{-0.16}$   &4890   &  resonance states  \\  \hline
$\mathcal{J}_{2,\mu}^+$     &$1^-$                        &$5.80_{-0.08}^{+0.09}$  &$0.54^{+0.09}_{-0.08}$   &4920              &  resonance states  \\      \hline
$\mathcal{J}_{3,\mu}^+$     &$1^-$    &$5.71_{-0.07}^{+0.07}$  &$0.86^{+0.13}_{-0.12}$   &4920   &   resonance states  \\
\hline
\hline
\end{tabular}
\end{center}
\end{table}

\section{Conclusions}
In the present work, twelve currents with the positive and negative charge conjugations, the partial and covariant derivatives are constructed to study the $P-wave$ $\bar{\Lambda}_c\Lambda_c$ dibaryon states in a consistent way. Analytical calculation of the spectral densities at the QCD sides show that the positive and negative charge conjugations should not have too much difference, and similar situations are there for the currents with the partial and covariant derivatives. The numerical results support these judgements. The masses of the $P-wave$  $\bar{\Lambda}_c\Lambda_c$ states are derived, they are all high above the corresponding thresholds of the two $\bar{\Lambda}_c$ and $\Lambda_c$ constituents. The present study show that it is unlikely for the $\bar{\Lambda}_c\Lambda_c$ with $P-wave$  to form the bound hadronic molecule, they should be the resonance dibaryon states.
\section*{Acknowledgements}
This work is supported by the National Natural Science Foundation with Grant Number 12575083 and the Natural Science Foundation of HeBei Province with Grant Number A2024502002.

\section*{Appendix}
The spectral density $\rho_{1,{\rm QCD}}^+(s)$ of the current $\mathcal{J}^{+}_1(x)$ is expressed three types of integrals,
\begin{eqnarray}
\notag \rho_{1,{\rm QCD}}^+(s)&=&\sum\limits_{\mathcal{A},\mathcal{B},\mathcal{C},n,i}\rho_{\mathcal{A},\mathcal{B},\mathcal{C}}(s,n)_i\,,
\end{eqnarray}
where, $\mathcal{A}$, $\mathcal{B}$ and $\mathcal{C}$ refer to the three types of integrals, $n$ is the dimension of the condensate, $i$ represents different kinds of vacuum condensates sharing the same dimension $n$. For the following expressions, $\tilde{m}^2=\frac{m_c^2}{(1-y)y}$ and $\bar{m}^2=\frac{(y+z)m_c^2}{y z}$, $y$ and $z$ are the parameters introduced during the calculation of the integrals in the momentum space of the two heavy quarks. For simplicity, set $\int_{y_i}^{y_f}dy\int_{z_i}^{1-y}dz\rightarrow\int dydz$
and $\int_{y_i}^{y_f}dy\rightarrow\int dy\,$. For type $\mathcal{A}$ and $\mathcal{B}$, $y_i=\frac{1}{2}(1-\sqrt{1-4m_c^2/s})$, $y_f=\frac{1}{2}(1+\sqrt{1-4m_c^2/s})$ and $z_i=\frac{y m_c^2}{y s-m_c^2}$. For type $\mathcal{C}$, $y_i=0$ and $y_f=1$.
\begin{eqnarray}
\notag \rho_{\mathcal{A}}(s,14) &=& - \frac{11 g_s^2 \langle\bar{q}q\rangle^3 \langle\bar{q}g_s\sigma Gq\rangle}{2592 \pi^4} \int dy \, (1-y) y \, (3s-2\tilde{m}^2)\,{\rm exp}\left(-\frac{s}{T^2}\right)\\
\notag && - \frac{m_c^2 \langle\bar{q}q\rangle^3 \langle\bar{q}g_s\sigma Gq\rangle}{24 \pi^2} \int dy\,{\rm exp}\left(-\frac{s}{T^2}\right) \,,
\end{eqnarray}
\begin{eqnarray}
\notag \rho_{\mathcal{A}}(s,16)_1 &=& \frac{g_s^2 \langle\bar{q}q\rangle^2 \langle\bar{q}g_s\sigma Gq\rangle^2}{384 \pi^4} \int dy\, (1-y) y\,{\rm exp}\left(-\frac{s}{T^2}\right)\\
\notag  && - \frac{\langle\bar{q}q\rangle^2 \langle\bar{q}g_s\sigma Gq\rangle^2}{32 \pi^2} \int dy\, (1-y) y\,{\rm exp}\left(-\frac{s}{T^2}\right)\,,
\end{eqnarray}
\begin{eqnarray}
\notag \rho_{\mathcal{A}}(s,16)_2 &=& \frac{\langle g_s^2GG\rangle \langle\bar{q}q\rangle^4}{144 \pi^2} \int dy\, (1-y) y\,{\rm exp}\left(-\frac{s}{T^2}\right)\,,
\end{eqnarray}
\begin{eqnarray}
\notag \rho_{\mathcal{B}}(s,0) &=& - \frac{m_c^2}{68812800 \pi^{10}} \int dy dz\, (1-y-z)^6 (s-\bar{m}^2)^7 \,{\rm exp}\left(-\frac{s}{T^2}\right) \\
\notag &&+ \frac{1}{137625600 \pi^{10}} \int dy dz\, y z (1-y-z)^6 (s-\bar{m}^2)^7 (5s-\bar{m}^2) \,{\rm exp}\left(-\frac{s}{T^2}\right)\,,
\end{eqnarray}
\begin{eqnarray}
\notag \rho_{\mathcal{B}}(s,4) &=& \frac{m_c^2 \langle g_s^2GG\rangle}{58982400 \pi^{10}} \int dy dz \,\frac{(1-y-z)^6}{y^2} (s-\bar{m}^2)^4 [5s y -2(1-y) (s-\bar{m}^2)]\,{\rm exp}\left(-\frac{s}{T^2}\right) \\
\notag && - \frac{m_c^2 \langle g_s^2GG\rangle}{29491200 \pi^{10}} \int dy dz\, \frac{z (1-y-z)^6}{y^2} (s-\bar{m}^2)^4 (7s-2\bar{m}^2) \,{\rm exp}\left(-\frac{s}{T^2}\right) \\
\notag && - \frac{m_c^2 \langle g_s^2GG\rangle}{13107200 \pi^{10}} \int dy dz\, \frac{(1-y-z)^5}{y} (s-\bar{m}^2)^5 \,{\rm exp}\left(-\frac{s}{T^2}\right) \\
\notag && - \frac{m_c^2 \langle g_s^2GG\rangle}{3932160 \pi^{10}} \int dy dz\, (1-y-z)^4 (s-\bar{m}^2)^5 \,{\rm exp}\left(-\frac{s}{T^2}\right) \\
\notag && + \frac{\langle g_s^2GG\rangle}{39321600 \pi^{10}} \int dy dz\, z (1-y-z)^5 (s-\bar{m}^2)^5 (4s-\bar{m}^2) \,{\rm exp}\left(-\frac{s}{T^2}\right) \\
\notag && + \frac{\langle g_s^2GG\rangle}{3932160 \pi^{10}} \int dy dz\, y z (1-y-z)^4 (s-\bar{m}^2)^5 (4s-\bar{m}^2) \,{\rm exp}\left(-\frac{s}{T^2}\right)\,,
\end{eqnarray}

\begin{eqnarray}
\notag \rho_{\mathcal{B}}(s,6) &=& \frac{g_s^2 m_c^2 \langle\bar{q}q\rangle^2}{331776 \pi^8}\int dy dz\,(1-y-z)^3 (s-\bar{m}^2)^4 \,{\rm exp}\left(-\frac{s}{T^2}\right) \\
\notag && + \frac{19 g_s^2 \langle\bar{q}q\rangle^2}{1658880 \pi^8} \int dy dz\, y z(1-y-z)^3 (s-\bar{m}^2)^4 (7s-2\bar{m}^2)\,{\rm exp}\left(-\frac{s}{T^2}\right) \\
\notag && - \frac{m_c^2 \langle\bar{q}q\rangle^2}{1536 \pi^6} \int dy dz\,(1-y-z)^3 (s-\bar{m}^2)^4 \,{\rm exp}\left(-\frac{s}{T^2}\right) \\
\notag && + \frac{\langle\bar{q}q\rangle^2}{3840 \pi^6} \int dy dz\, y z (1-y-z)^3 (s-\bar{m}^2)^4 (7s-2\bar{m}^2) \,{\rm exp}\left(-\frac{s}{T^2}\right) \,,
\end{eqnarray}

\begin{eqnarray}
\notag \rho_{\mathcal{B}}(s,8) &=& \frac{m_c^2 \langle\bar{q}q\rangle \langle\bar{q}g_s\sigma Gq\rangle}{512 \pi^6} \int dy dz\, (1-y-z)^2 (s-\bar{m}^2)^3 \,{\rm exp}\left(-\frac{s}{T^2}\right) \\
\notag &&- \frac{\langle\bar{q}q\rangle \langle\bar{q}g_s\sigma Gq\rangle}{256 \pi^6} \int dy dz\, y z (1-y-z)^2 (s-\bar{m}^2)^3 (3s-\bar{m}^2) \,{\rm exp}\left(-\frac{s}{T^2}\right)\,,
\end{eqnarray}
\begin{eqnarray}
\notag \rho_{\mathcal{B}}(s,10)_1 &=& - \frac{575 m_c^2 \langle\bar{q}g_s\sigma Gq\rangle^2}{393216 \pi^6} \int dy dz\, (1-y-z)   (s-\bar{m}^2)^2 \,{\rm exp}\left(-\frac{s}{T^2}\right) \\
\notag && - \frac{49 m_c^2 \langle\bar{q}g_s\sigma Gq\rangle^2}{98304 \pi^6} \int dy dz \, \frac{(1-y-z)^2}{y} (s-\bar{m}^2)^2 \,{\rm exp}\left(-\frac{s}{T^2}\right) \\
\notag &&+ \frac{\langle\bar{q}g_s\sigma Gq\rangle^2}{2048 \pi^6} \int dy dz\, y z(1-y-z) (s-\bar{m}^2)^2 (15s-6\bar{m}^2) \,{\rm exp}\left(-\frac{s}{T^2}\right) \\
\notag && + \frac{\langle\bar{q}g_s\sigma Gq\rangle^2}{98304 \pi^6} \int dy dz\, z (1-y-z)^2 (s-\bar{m}^2)^2 (5s-2\bar{m}^2) \,{\rm exp}\left(-\frac{s}{T^2}\right) \\
\notag && - \frac{\langle\bar{q}g_s\sigma Gq\rangle^2}{65536 \pi^6} \int dy dz\,(1-y-z)^3 (s-\bar{m}^2)^2 (5s-2\bar{m}^2) \,{\rm exp}\left(-\frac{s}{T^2}\right) \,,
\end{eqnarray}
\begin{eqnarray}
\notag  \rho_{\mathcal{B}}(s,10)_2&=& - \frac{m_c^2 \langle g_s^2GG\rangle \langle\bar{q}q\rangle^2}{2304 \pi^6} \int dy dz \, \frac{(1-y-z)^3}{y^2} (s-\bar{m}^2) [(1-y) (s-\bar{m}^2) -s y ]\,{\rm exp}\left(-\frac{s}{T^2}\right) \\
\notag && - \frac{m_c^2 \langle g_s^2GG\rangle \langle\bar{q}q\rangle^2}{1152 \pi^6} \int dy dz \, \frac{z(1-y-z)^3}{y^2}   (s-\bar{m}^2) (2s-\bar{m}^2) \,{\rm exp}\left(-\frac{s}{T^2}\right) \\
\notag && - \frac{m_c^2 \langle g_s^2GG\rangle \langle\bar{q}q\rangle^2}{2048 \pi^6} \int dy dz \, \frac{(1-y-z)^2}{y} (s-\bar{m}^2)^2 \,{\rm exp}\left(-\frac{s}{T^2}\right) \\
\notag && + \frac{m_c^2 \langle g_s^2GG\rangle \langle\bar{q}q\rangle^2}{1536 \pi^6} \int dy dz\,(1-y-z) (s-\bar{m}^2)^2 \,{\rm exp}\left(-\frac{s}{T^2}\right) \\
\notag && + \frac{\langle g_s^2GG\rangle \langle\bar{q}q\rangle^2}{6144 \pi^6} \int dy dz\, z (1-y-z)^2 (s-\bar{m}^2)^2 (5s-2\bar{m}^2) \,{\rm exp}\left(-\frac{s}{T^2}\right) \\
\notag && + \frac{5\langle g_s^2GG\rangle \langle\bar{q}q\rangle^2}{9216 \pi^6} \int dy dz\, y z(1-y-z) (s-\bar{m}^2)^2 (5s-2\bar{m}^2)\,{\rm exp}\left(-\frac{s}{T^2}\right) \,,
\end{eqnarray}

\begin{eqnarray}
\notag \rho_{\mathcal{B}}(s,12) &=& \frac{5 g_s^2 m_c^2 \langle\bar{q}q\rangle^4}{1296 \pi^4} \int dy dz \,(s-\bar{m}^2) \,{\rm exp}\left(-\frac{s}{T^2}\right) \\
\notag && + \frac{11 g_s^2 \langle\bar{q}q\rangle^4}{1296 \pi^4} \int dy dz\, y z (s-\bar{m}^2) (2s-\bar{m}^2)\,{\rm exp}\left(-\frac{s}{T^2}\right) \,,
\end{eqnarray}

\begin{eqnarray}
\notag \rho_{\mathcal{C}}(s,14) &=& - \frac{7 g_s^2 m_c^2 \langle\bar{q}q\rangle^3 \langle\bar{q}g_s\sigma Gq\rangle}{2592 \pi^4} \int dy \,\delta(s-\tilde{m}^2)\,{\rm exp}\left(-\frac{s}{T^2}\right)\,,
\end{eqnarray}
\begin{eqnarray}
\notag \rho_{\mathcal{C}}(s,16)_1 &=& \frac{7 g_s^2 m_c^2 \langle\bar{q}q\rangle^2 \langle\bar{q}g_s\sigma Gq\rangle^2}{20736 \pi^4} \int dy \, \delta(s-\tilde{m}^2)  \left(\frac{s}{T^2}+1\right)\,{\rm exp}\left(-\frac{s}{T^2}\right) \\
\notag && + \frac{g_s^2 \langle\bar{q}q\rangle^2 \langle\bar{q}g_s\sigma Gq\rangle^2}{2304 \pi^4} \int dy \, \delta(s-\tilde{m}^2) \, (1-y) y \left(\frac{s^2}{T^2}+4s\right)\,{\rm exp}\left(-\frac{s}{T^2}\right) \\
\notag && + \frac{5 m_c^2 \langle\bar{q}q\rangle^2 \langle\bar{q}g_s\sigma Gq\rangle^2}{192 \pi^2}  \int dy \, \delta(s-\tilde{m}^2) \left(\frac{s}{T^2}+1\right)\,{\rm exp}\left(-\frac{s}{T^2}\right) \\
\notag && - \frac{m_c^2 \langle\bar{q}q\rangle^2 \langle\bar{q}g_s\sigma Gq\rangle^2}{288 \pi^2} \int dy \, \delta(s-\tilde{m}^2)  \frac{1}{y} \,{\rm exp}\left(-\frac{s}{T^2}\right) \\
\notag && - \frac{\langle\bar{q}q\rangle^2 \langle\bar{q}g_s\sigma Gq\rangle^2}{192 \pi^2} \int dy \, \delta(s-\tilde{m}^2)\,  (1-y) y \left(\frac{s^2}{T^2}+4s\right)\,{\rm exp}\left(-\frac{s}{T^2}\right) \,,
\end{eqnarray}
\begin{eqnarray}
\notag \rho_{\mathcal{C}}(s,16)_2 &=& \frac{m_c^2 \langle g_s^2GG\rangle \langle\bar{q}q\rangle^4}{288 \pi^2} \int dy \, \delta(s-\tilde{m}^2)  \left(\frac{s}{T^2}+1\right)\,{\rm exp}\left(-\frac{s}{T^2}\right) \\
\notag && + \frac{\langle g_s^2GG\rangle \langle\bar{q}q\rangle^4}{864 \pi^2} \int dy \, \delta(s-\tilde{m}^2) \, (1-y) y \left(\frac{s^2}{T^2}+4s\right)\,{\rm exp}\left(-\frac{s}{T^2}\right)\,.
\end{eqnarray}

\end{document}